\documentclass[12pt,twoside,dvips]{article}
\usepackage{epsfig}
\usepackage{a4wide}

\newcommand{\bmath}[1]{\mbox{\boldmath${#1}$}}

\begin{document}
\mbox{ }
\vspace{20mm}

\includegraphics{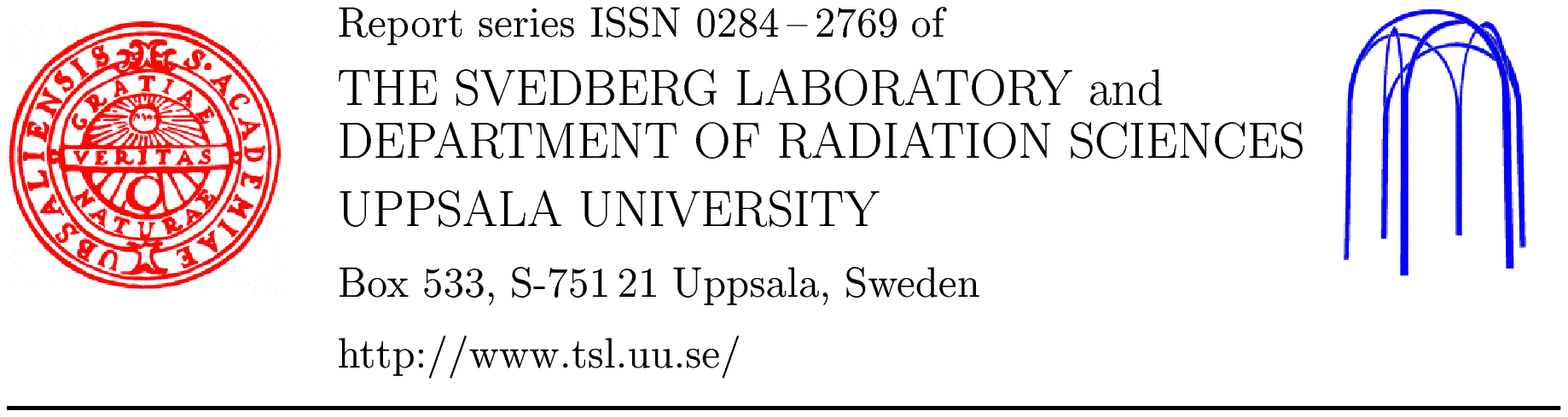}

\begin{flushright}
\begin{minipage}[t]{37mm}
{\bf TSL/ISV-99-0221 \\
November 1999} 
\end{minipage}
\end{flushright}

\begin{center}
\vspace{20mm}

{\bf \Large
Multi-pion production in the \boldmath$dd\rightarrow\alpha X$ reaction}

\vspace{15mm}

{\Large Anders G{\aa}rdestig\footnote{Electronic address: grdstg@tsl.uu.se}}\\
\vspace{3mm}
Nuclear Physics Division, Uppsala University,
Box 535, S-751 21 Uppsala, Sweden\\

\end{center}

\vspace{15mm}

{\bf Abstract:} 
A simple model, based on two parallel and independent $N\!N\to d\pi$
processes, has recently been proposed for two-pion production in the
$dd\to\alpha X$ reaction. It reproduces all observed features,
including the sharp peak structure in momentum distributions (the ABC
effect) and the strong oscillations in the deuteron vector and tensor
analyzing powers. This model is now extended to describe also
four-pion production with the same basic mechanism, but with two
$np\to d\pi\pi$ processes as input. The calculations of the high
missing mass spectra are within about 30\% of the experimental data
for beam energies in the range $1.9<T_d<2.4$~GeV.

\vfill
\clearpage
\setcounter{page}{1}

It has long been known that the momentum distributions of the $np\to
dX$~\cite{HH,Ban71,Plouin}, $pd\to~^3{\rm He}X$~\cite{ABC,Ban73}, and
$dd\to\alpha X$~\cite{Ban76,SPESIII} reactions at intermediate
energies share some common and spectacular features. Characteristic
are the sharp peaks, located slightly above the two-pion threshold, in
association with a broader bump around the maximal missing mass.  This
behaviour, known as the ABC effect, is observed only in isospin
$I_X=0$ channels. The phenomenon is believed to originate from a
largely kinematic enhancement in the production of two pions, but all
previous attempts to give a quantitative description have so far had
only limited success. In particular, the theoretical models for the
$np\to dX$~\cite{BRS,ALS76,Barry,KE} and
$pd\to~^3{\rm He}X$~\cite{ALS73,Barry} reactions have problems in
reproducing the angular dependences~\cite{Plouin,ALS73}. A recently
published model for $dd\to\alpha X$~\cite{GFW2} has, however, had
remarkable success in reproducing data. It is based on two almost free
$N\!N\to d\pi$ processes happening in parallel, where the two final
deuterons merge to form the $\alpha$ particle.  Since the deuteron is
loosely bound, these processes are almost free and their c.m.\ systems
coincide with the overall c.m.\ frame. Because of the strong
forward-backward peaking in the two $d\pi$ systems, the small missing
mass (parallel pions) and maximal missing mass (back-to-back pions)
configurations are both enhanced. In the latter case, the large
relative momentum between the deuterons in the $\alpha$ particle does
cause a suppression relative to the low $\pi\pi$ mass case. The ABC
peaks and the central bump are hence simultaneously explained. The
model, depicted by the Feynman diagram in Fig.~1, is able to reproduce
quantitatively all the observed features of $dd\to\alpha X$ in the
energy range $0.8 < T_d < 1.9$~GeV, including angular and energy
distributions~\cite{GFW2} and deuteron vector and tensor analyzing
powers~\cite{SPESIII,GFW2}. The purpose of this work is to use the
same type of mechanism to estimate four-pion production with two
parallel $np\to d\pi\pi$ processes as input.

At higher energies ($T_d > 1.9$~GeV), where production of heavier
mesons and multiple ($>2$) pions begins to be important, the central
bump in $2\pi$-production at large $\pi\pi$ effective masses is
strongly suppressed. The large relative momentum between the two pions
requires the deuterons to be far out in the tail of the
$\alpha$-particle wave function in momentum space. The angular
dependences of the subprocesses are here of minor importance --- it is
the dynamics of the $dd\!:\!\alpha$ vertex that dominates the
behavior. The situation will be very similar when heavier mesons are
produced by the same mechanism, the effect being more pronounced the
larger the maximal missing mass is compared to the meson mass. At high
enough energy the formation of the $\alpha$ particle will force the
two mesons to be close together in momentum space and suppress other
configurations. For the production of pairs of mesons with definite
masses we then expect sharp peaks close to the two-meson threshold and
a rapid decrease towards maximal missing mass.  On the other hand,
when the meson has a broad width, like the controversial $\sigma$
meson, the peaks corresponding to different masses should be added
together with appropriate weights. In the present analysis we will use
the $\sigma$ meson as a convenient tool to describe and handle the
$I_X=0$ pion pair produced in $np\to dX$. By considering two parallel
$\sigma$ productions, the four-pion contribution could be estimated in
the same framework as the two-pion one. The missing mass distributions
of $np\to dX$ is then the weight functions needed as input to this
$dd\to\alpha\sigma\sigma$ model.

Since the mathematical details of the model have been published~\cite{GFW2}, 
only the main features are recapitulated here. The dashed lines in the 
Feynman diagram (Fig.~1) are now considered to be $\sigma$ mesons with the 
(possibly unequal) masses $m_1$ and $m_2$. Instead of the complicated and many 
$N\!N\to d\pi$ amplitudes used in our previous work, we here assume that the 
$np\to \sigma d$ reaction can be parametrized by one single $s$-wave amplitude 
\begin{equation}
    \mathcal{M}_{np\to \sigma d} = \mathcal{A}\,\eta_n^{c\dagger}
    \left[ -\frac{1}{\sqrt{2}}\bmath{\sigma\cdot\epsilon}^{\dagger}_d 
    \right]\eta_p,
\end{equation}
where $\eta_x$ is a nonrelativistic two-component spinor,
$\eta^c=-i\sigma_2\eta^{\ast}$ its charge conjugate,
$\bmath{\epsilon}_d$ the deuteron polarization vector, and $\mathcal{A} =
\mathcal{A}(m_{\sigma})$ the $\sigma$ meson production amplitude. After
summing over internal spins the spin kernel $\mathcal{K}$~\cite{GFW2} is,
with this normalization,
\begin{equation}
    \mathcal{K} = \frac{\bmath{\epsilon_1\cdot\epsilon_2}}{2\sqrt{3}} 
    \mathcal{A}^2,
\end{equation}
where $\bmath{\epsilon}_{i}$ are the polarization vectors of the initial
deuterons. The summed square is then $\sum|\mathcal{K}|^2=(1/4)|\mathcal{A}|^4$,
but an extra factor of four is needed because there are two ways of distributing
the four nucleons between the two $\sigma d$ vertices.

The mass dependence of the amplitude $\mathcal{A}$ was obtained from 
experimental data for $np\to dX$~\cite{HH,Plouin} by transforming the laboratory
momentum distribution to a missing mass distribution in the c.m.\
frame, where it should be symmetric. 
The data were taken with a spectrum of neutron momenta and their
quality suggests that one should use the forward ABC peak (low 
$k_{\alpha}^{{\rm lab}}$) at the highest energy~\cite{Plouin}, while the
backward ABC is the better choice at 991~MeV~\cite{HH}. By integrating this 
over a small mass range $\Delta m_x$ the differential cross section for 
$\sigma d$ could be estimated:
\begin{equation}
    \left.\frac{d\sigma}{d\Omega}(np\to \sigma d)\right|_{{\rm cm}} \approx 
    \frac{d^2\sigma}{d\Omega dm_x}\Delta m_x.
\label{eq:dsig}
\end{equation}
In the normalization we are using, the relation between the cross
section and the amplitude, and in
particular its mass dependence, is given by
\begin{equation}
    \left.\frac{d\sigma}{d\Omega}(np\to \sigma d)\right|_{{\rm cm}}
    \approx
    \frac{3}{16(2\pi)^2}\frac{m^2}{s_{\sigma d}}\frac{k}{p}\,
    \,\frac{d|\mathcal{A}|^2}{dm_x}\Delta m_x,
\label{eq:ampl}
\end{equation}
where $m$ is the nucleon mass and $p$ and $k$ are the nucleon and
deuteron c.m.\ momenta. The squared amplitude $d|\mathcal{A}|^2/dm_x$,
extracted from Eqs.~(\ref{eq:dsig}) and (\ref{eq:ampl}), was fitted to
a third degree polynomial in $m_x$ for each beam energy. At
$T_d=1163$~MeV an angular distribution has been measured~\cite{Plouin}
and the transformation was done for each of the angles. The variation
of the amplitude with the $\sigma$ mass is roughly independent of the
laboratory angle, which supports our $s$-wave assumption, and the
polynomial was fitted to all of the 1163~MeV data
simultaneously. Since none of the energies where the amplitude
functions were extracted coincides with the energies required for our
$dd$ estimate, a linear interpolation in ${\displaystyle
E_i=\sqrt{s_{\sigma_i d}}}$ was applied between the fitted curves. The
masses used in the fitted polynomials were estimated by letting them
be in the same proportion to the mass range at the fitted energies as
the actual mass is to its allowed mass range. The amplitudes were then
finally evaluated for each individual $m_x$ by multiplying the fitted
and interpolated values by the increment $\Delta m_x$.

The matrix element $\mathcal{M}_{\alpha\sigma\sigma}$ is written as the
product $\mathcal{M}=-i(m_{\alpha}/v_d)\mathcal{KW}$, where $\mathcal{W}$ is the
dimensionless form factor defined in Ref.~\cite{GFW2}. In this
factorization approximation the amplitudes are calculated without
considering the Fermi momenta. Nevertheless it is possible to get a
crude estimate of their influence by assuming that the distribution of
energy between the two subprocesses follows a simple Gaussian,
$\exp[-(\Delta E_i)^2/(2\sigma_E^2)]$. Here $\Delta
E_i=E_i-\sqrt{s_{\alpha\sigma\sigma}}/2$ is the deviation in energy
from equal sharing and $\sigma_E\approx25$~MeV is calculated from a
Gaussian deuteron wave function. The calculations are very stable
against variations of this distribution. Only for small assumed widths
($\sigma_E<5$~MeV) could significant changes be observed.

The calculation of the form factor $\mathcal{W}$ follows the lines
of~\cite{GFW2}, apart from obvious mass and kinematical changes. The
integration over meson angles was performed directly since there are
no angular or momentum dependences in the amplitudes. For each choice
of total missing mass $M_X$, the momentum distributions were
calculated for all possible combination of the two $\sigma$
masses. The results were then accumulated to give the full four-pion
spectrum. With this procedure, all the allowed four-body pion phase
space was completely accounted for.

Our predictions of the four-pion missing mass distributions in the
c.m.\ frame can be seen in Fig.~2 together with data points taken
from~\cite{Ban76}. Note that the scale factors applied to the
calculations are not directly comparable to those in~\cite{GFW2}; the
latter values should be multiplied by the relativistic factor
$1/\gamma^2=(m_d/E_d)^2$, which was overlooked in the programs (but not
in the formulae) of that reference.

A plausible reason for the underestimate of the cross section is the
neglect of contributions from $pp\to pp\pi\pi$. The cross section for
this process is at $\approx1.2$~GeV about half that for $np\to
d\pi\pi$~\cite{Shim} but, lacking information about angular
distributions and the $nnpp\!:\!\alpha$ wave function, it is
impossible to describe its influence.  Another effect will occur at
low missing masses ($M_X\approx 4m_{\pi}$), where the statistics of 4
equal bosons should increase the counting rate. In the semi-classical
model we are employing --- and with no knowledge of the $np\to
d\pi\pi$ amplitudes --- this effect cannot be estimated.  Distortion
due to rescattering of the pions might decrease the cross section.

Superposed upon all three of the $4\pi$ spectra there is an additional
structure which, since it peaks at $M_X\approx 780$~MeV/$c^2$
independent of beam energy, might be attributed to the $\omega$
meson. This interpretation is however not unambiguous since the widths
of the peaks are much larger than the stated experimental
resolution. The experimentalists noted this difference and remarked
that the resolution was calculated from known properties of the beam
and spectrometer but that they were never able to test it directly
because of poor statistics for two-body reactions in $dd\to\alpha
X$~\cite{LeBrun}. The upper curves in Fig.~2 result from single
Gaussians fitted to these points which, after integration give the
c.m.\ cross sections $d\sigma/d\Omega(\mbox{`$\omega$'})=$~9, 6, and
2~nb/sr for 1.9, 2.2, and 2.4~GeV. Because of the threshold $4\pi$
probable enhancement mentioned in the previous paragraph, these values
are probably overestimates.

The assumption that the four pions are produced via two $\sigma$'s has
the consequence that the relative weights for charged and neutral
pions should be $N(++--):N(0\,0+-):N(0\,0\,0\,0)=4\!:\!4\!:\!1$, which
would be a test of the model, since other possible production channels
have different weights.  In the case of $\rho\rho$ production, where
$4\pi^0$ is forbidden, $N(++--):N(0\,0+-)=2\!:\!1$. This contribution
can be estimated from $pp\to d\pi^0\pi^+$. The cross section for this
process is smaller than that for $np\to d\pi\pi$ by roughly a factor
of 10~\cite{Shim} and, in addition, there is only one diagram in this
case, giving an extra factor 1/4. In total this will give a
contribution of less than 1\% of the $\sigma\sigma$ cross section.

Despite the crudeness and simplistic nature of the assumptions and
approximations made in this model, the calculations are in pleasing
agreement with data. Taken together with the results of \cite{GFW2},
it seems likely that most of the multi-meson production in the
$dd\to\alpha X$ reaction in the $0.8<T_d<2.4$~GeV range could be
attributed to the same general double-interaction mechanism.  Further
support for this suggestion could be obtained from the estimate of
$\gamma\gamma$ production via two $np\to d\gamma$ processes. This is
shown to be of considerable importance as background to the
charge-symmetry-breaking $dd\to\alpha\pi^0$ reaction~\cite{DFGW}.

I am grateful to Colin Wilkin for suggesting this approach and would like to
thank him and G\"oran F\"aldt for stimulating discussions of the results.

\noindent
\begin{figure}[p]
\vspace{1cm}
\includegraphics{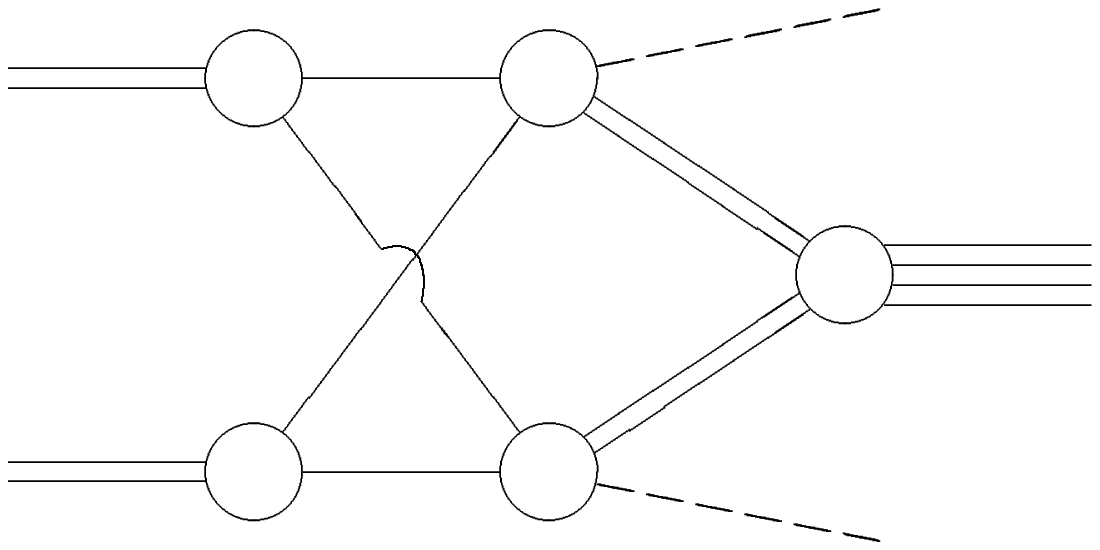}
\caption{The Feynman diagram for the $dd\to\alpha\pi\pi$ and
$dd\to\alpha\sigma\sigma$ reactions.}
\end{figure}
\begin{figure}[p]
\includegraphics{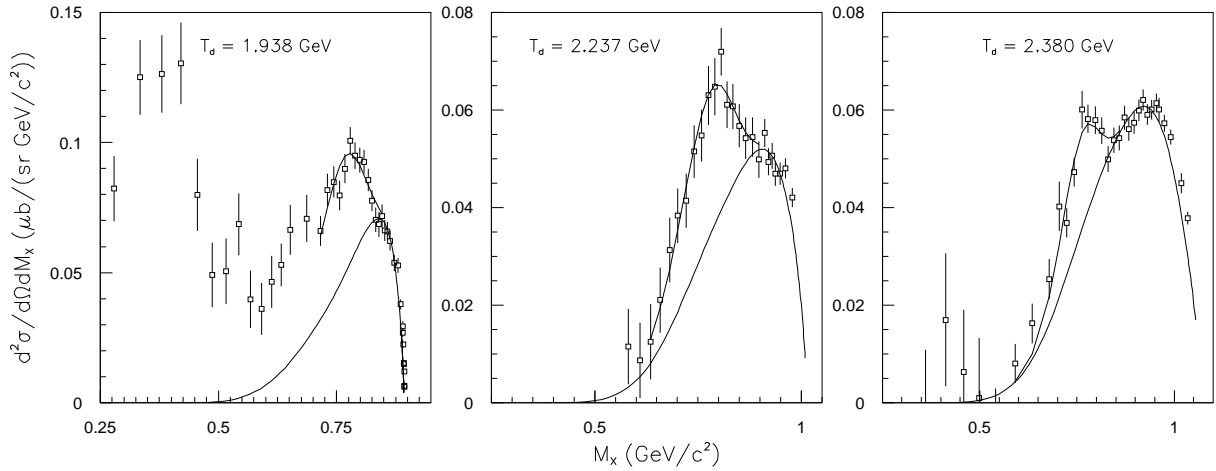}
\caption{The missing mass distributions for $dd\to\alpha X$ at
$\theta_{\alpha}^{{\rm cm}}\approx 0^{\circ}$. The data~\protect\cite{Ban76}
are compared with the $\sigma\sigma$ (lower solid line) model, the latter curves
multiplied by scale factors 1.2, 1.0, and 1.3 for 1.9, 2.2, and 2.4~GeV. The 
upper solid line is the fit to the `$\omega$' peak.}
\end{figure}

\end{document}